\documentstyle[epsfig]{aipproc}

\begin{document}
\title{Kinematic Arguments Against Single\\
Relativistic Shell Models for GRBs}

\author{E. E. Fenimore, E. Ramirez, and M. C. Sumner}
\address{
NIS-2, MS D436, Los Alamos National Laboratory, Los Alamos NM 87545}

\maketitle

\begin{abstract}
Two main types of models have been suggested to explain the long durations
and multiple peaks of Gamma-Ray Bursts (GRBs).
In one, there is a very quick release of
energy at a central site resulting in a single relativistic shell that
produces peaks in the time history through its interactions with the
ambient material.  In the other, the central site sporadically releases
energy over hundreds of seconds forming a peak with each burst of energy.
We show that the average envelope of emission and the presence of gaps in
GRBs are inconsistent with a single relativistic shell.  We estimate that
the maximum fraction of a single shell that can produce gamma-rays in a
GRB with multiple peaks is $10^{-3}$, implying that single
relativistic shells require $10^3$ times more energy than previously
thought.  We conclude that either the central site of a GRB must produce
$\sim 10^{51}$ erg s$^{-1}$ for hundreds of seconds, or the
relativistic shell must have structure on a scales the order of
$\sqrt\epsilon\Gamma^{-1}$, where $\Gamma$ is the
bulk Lorentz factor ($\sim 10^2$ to
$10^3$) and $\epsilon$ is the efficiency.
\end{abstract}
\section*{Introduction}
Two classes of models have arisen that explain different (but not all) 
aspects of the duration of GRBs.
In the ``external'' shock model\cite{meszrees93},
the release of energy is very quick and a relativistic
shell forms that expands outward for a long period of time ($10^5$ to $10^7$
sec).
At some point, interactions with the external medium
(hence the name)
cause the energy of the bulk motion to be converted to gamma-rays.
The alternative theory is that a central site releases energy in the form
of a wind or multiple shells over a period of time commensurate with the
observed duration of the GRB\cite{reesmesz94}.
The gamma-rays are produced by the 
internal interactions within the wind, hence these scenarios are often 
referred to as internal shock models.

In Fenimore, Madras, \& Nayakshin\cite{FMN96}, we used kinematics
to demonstrate that a
single relativistic shell has extreme difficulties
explaining the observed GRB time structure.  We have made direct
comparisons to the observations for three of the most potent
arguments: the average envelope\cite{FS97,F98}, gaps in the time
history\cite{FMN96}, and the maximum active fraction of the
shell\cite{FCRS98}.  In this paper, we summarize those arguments.
\subsection*{Argument 1: Average Envelope}
If a single relativistic shell with high bulk Lorentz factor ($\Gamma$)
expands outward from a central site towards an observer, the observed
time structure is dominanted by two effects.  First,
although the shell might produce gamma-rays for a long period of time
(say $t_0$ to $t_{\rm max}$),
the shell keeps up with the photons
such that they arrive at a detector over a short period of time.
If the shell has velocity $v = \beta c$ such that the Lorentz factor,
$\Gamma$ is $(1-\beta^2)^{-1/2}$, then photons emitted over a period $t$
arrive at a detector over a much shorter period, $T = (1-\beta)t \approx
t/(2\Gamma^2)$.
Second, the curvature causes regions of the shell off-axis to arrive
later at the detector. The additional distance that photons must travel
is $\sim R(1-\cos\theta)$ where R is the radius
of the shell ($\sim ct$).
At a typical observable angle of $\theta = \Gamma^{-1}$,
the delay due to the curvature is the same order as the time scale
of arrive for on-axis photons:
$t/(2\Gamma^2)$.
In \cite{FMN96,FS97}, we showed that a single symmetric shell produces
a  ``FRED''-like shape (fast rise, rapid decay):
\begin{eqnarray}
V(T) &=&
V_0
{T^{\omega} - T_{0}^{\omega} \over T^{\alpha+1}}
~~~~~~~~~~~{\rm if}~T_{0} < T < T_{\rm max}\nonumber\\
&=&
V_0
{T_{\rm max}^{\omega} - T_{0}^{\omega} \over  T^{\alpha+1}}
~~~~~~~~~~~~~~~~{\rm if}~T > T_{\rm max}
\end{eqnarray}
where $V_0$ is a constant, $\omega = \alpha +3 -\nu$,
$\alpha$ is the spectral number index (e.g., 1.5),
and $\nu$ is a power law index for the intrinsic variation of the
shell's emissivity as a function of time.
The expansion effects occur in the rise of the envelope and
the curvature dominates in the fall.
We have also shown\cite{FS97} that during the decay phase,
the spectra should
evolve as $T^{-1}$.

\begin{figure}[t]
\centerline{\epsfig{file=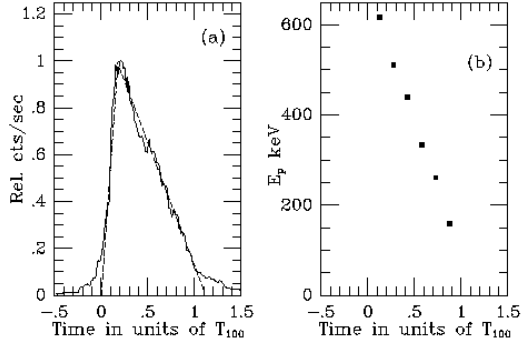,height=1.875in,clip=}}
\caption{
The average temporal and spectral evolution of bright events with
intermediate durations ($T_{90}$ between 16 and 40 s) based on the
BATSE MER data.
(a) The average time history.  The decay phase starting 20\% after the
beginning of the $T_{100}$ period is linear rather than the expected
power law.
(b) The average evolution of the
the peak of the $\nu F_\nu$ distribution.
The peak energy is also a linear function rather than the expected $T^{-1}$.
 These patterns
are inconsistent with that expected from
a single relativistic shell.
}
\label{fig1}
\end{figure}

  To test this, we have added together
32 bright BATSE bursts with durations between 16 and 40 s.
We align each burst by scaling it to a standard duration
defined to be $T_{100} =
(T_{90}+T_{50})/0.7$ where $T_{90}$ and $T_{50}$ are the durations
that contain 90\% and 50\% of the counts.
Figure \ref{fig1} is from
ref \cite{F98} which should be consulted for compete details.
The average envelope {\it and} the average spectral evolution are
linear whereas
a single relativistic shell
predicts that they should be power laws with indexes $-\alpha-1$ and -1,
respectively.  We conclude that the average envelope of GRBs
is not consistent with a single relativistic shell.
\subsection*{Argument 2: Gaps in Time History}
\begin{figure}[t]
\centerline{\epsfig{file=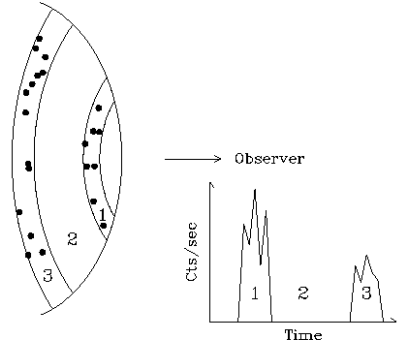,height=2.0in,clip=}}
\caption{
Schematic of the relationship between the emission on a shell and
the observed time history.
The
curvature delays the photons from off-axis regions such that at any
one time, the observer sees photons from an annulus oriented around
the line of sight.  The
perpendicular size of the shell is $\sim \Gamma T$ whereas a causally
connected entity (represented by the dots) is only $\Gamma\Delta T$.
Here, $T$ and $\Delta T$
are the time in the time history and a typical time scale of variation.
Gaps imply that entire causally disconnected regions
do not emit (e.g., region 2 produces gap 2 in the time history).
The number of entities in each annulus determines the
variability of the time history.
}
\label{fig2}
\end{figure}

Gaps or precursors in GRBs produce the strongest evidence against
a single
relativistic shell.
The sharp rise in the average profile indicates that the shell emits for
a short
period of time (i.e., $t_0$ to $t_{\rm max}$ is short
relative to the duration of the event), so that the
shape of the overall envelope is dominated by photons delayed by the
curvature.
During the decay phase, the region that can contribute
photons to a given section of the time history  is an
annulus oriented about the line of sight
to the observer (see Fig.
\ref{fig2}).
 Gaps in the time history indicate that some annuli
emit while others do not.  These annuli are causally disconnected,
making it difficult to achieve this large scale coherence.
  (See Figure 7 in
\cite{FMN96} for
attempts to fit the emission of shells to bursts with gaps and
precursors.)

\subsection*{Argument 3: Active Fraction of Shell}
\begin{figure}[t]
\centerline{\epsfig{file=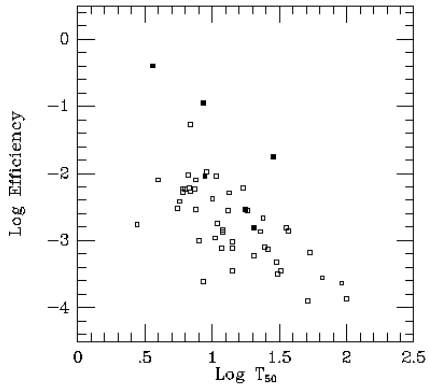,height=2.1in,clip=}}
\caption{
 Typical values of the fraction of a relativistic shell that becomes
active during a GRB as a function of the duration of the emission
($T_{50}$). 
The six solid squares are FRED-like BATSE bursts for
which direct estimates of the size of the shell can be made.  The 46 open 
squares are long complex BATSE bursts where we estimate the size in a manner
similar to the FRED-like estimate.
Under most conditions,
the efficiency is $\sim 0.1 \Delta T/T$.
These low values imply that either only a small fraction of
the shell converts its energy into gamma-rays or that GRBs consist
of very fine jets with angular sizes much smaller than
$\Gamma^{-1}$.
}
\label{fig3}
\end{figure}

Each dot in Fig. 2 is a causally connected region.
Note region 3 has more dots so it produces a
smoother time history (the intensity is less because the emission is
off axis so fewer photons are  beamed towards the observer).  We have
shown\cite{SF97} that the volume of the annulus that contributes at any
time is a constant so all sections of a time history should have about the
same smoothness.
We assume that the ``peaks'' in a time history represent Poisson
fluctuations in the number of entities contributing at any time.  We
determine the total number of entities ($N_N = \mu_N (T/\Delta T)$)
up to time $T$ by determining the rate of entities:
 $\mu_N = N^2/\delta N^2$ where $N$ and $\delta N$ are
the mean and root-mean-square of the profile. The fraction of the
shell that became active is $\epsilon = N_NA_N/A_S$. Here, $A_N$ is
the size of each entity ($=\pi c^2\Gamma^2\Delta T^2/k$, where $k$ is
13 for entities arising from entities that grow at the speed of sound and
is 1 for interactions with interstellar matter
(ISM) clouds, see \cite{FCRS98}).
The total area of the shell is
$A_S = 4\pi c^2\Gamma^2T^2f$ where $f$ is the fraction of the shell
out to $\theta = \Gamma^{-1}$ that contributes up to time $T$.  For
FRED-like bursts, Eq. 1 usually fits the profile such that $f$ is unity at
$T=0.8T_{50}$. For non-FRED like bursts, we simply assume that $f=1$
at $T=0.8T_{50}$.
Figure \ref{fig3} gives the efficiency for 6 FRED-like bursts and 46 bright,
long complex BATSE bursts based on
\begin{equation}
\epsilon =N_N \bigg[{\Delta T \over 2T}\bigg]^2 {1 \over kf}=
{N^2 \over (\delta N)^2} {\Delta T \over 3.2kT_{50}}
\end{equation} 
where we have used the case of shocks growing at the speed of sound.
For complete details see reference \cite{FCRS98}.
Thus, the spikiness of GRB time histories implies that only $\sim 10^{-3}$
of the surface of a shell becomes active.
This is lower than previously estimated\cite{FMN96,SP97}, and
implies that models require $\epsilon^{-1} \sim 10^3$ times more energy
than previously thought. Of course, reducing the fraction of the sky
into which
each shell expands can compensate for low efficiency for the
small price of requiring a higher density (by $\epsilon^{-1}$) of GRBs
in the universe.

 A common misconception is that one can just use ISM
clouds that cover most of the  shell's surface.
Each cloud could cause a relatively large peak while efficiently utilizing
the area of the shell.
This does not work
because the curvature of the expanding shell prevents the shell from
engaging the cloud instantaneously.  Rather, the portion of the shell
at $\theta \sim \Gamma^{-1}$ requires $R(1-\cos\theta)/v$ longer
before it reaches the cloud.  Even if the cloud happens to have a
concave shape such that the shell reaches the cloud simultaneously
over a wide range of angles, the resulting photons at $\theta \sim
\Gamma^{-1}$ must travel farther to the detector resulting in emission
that is delayed by $R(1-\cos\theta)/c$.
Since the speed is weakly dependent on $\Gamma$ or the
ambient material, there is no reason to believe
variations in the ambient material could cause  the
shell to develop into a plane wave oriented towards the observer 
such that the  photons produced by an interaction with an ISM
cloud or a shock would arrive as a short flare.
Only the instantaneous interaction of two plain parallel surfaces
oriented perpendicular to the line of sight can produce
a short peak from large surfaces.

These three arguments make a strong case against single,
symmetric relativistic
shells that undergo variations either due to shocks or interactions with
the ISM. There are two alternative explanations.  First, one can accept
the internal shock models\cite{reesmesz94}.
  These models have two weaknesses:
there is a concern that internal shocks are rather inefficient, and
the long, complex time history of a GRB
requires the central site to produce $10^{51}$ erg s$^{-1}$ for hundreds
of seconds.
Second, one can retain the quick energy release associated with the single
shell but
break the spherical symmetry of the shell by having the emitting material
confined to fine jets with angular width the order
of $\sim \sqrt{\epsilon}\Gamma^{-1}$.


\begin{references}
\def\Mesz{M\'esz\'aros\ }
\bibitem{meszrees93}\Mesz, P., and Rees, M. J., {\it ApJ}\ {\bf 405},
278 (1993).
\bibitem{reesmesz94}Rees, M. J., and \Mesz, P., {\it ApJ}\ {\bf 430},
L93 (1994).
\bibitem{FMN96}Fenimore, E. E., Madras, C. D., and Nayakshin, S.,
{\it ApJ}\ {\bf 473}, 998 (1996) astro-ph/9607163.
\bibitem{FS97}Fenimore, E. E., and Sumner, M. C.,
{\it All-Sky X-Ray Observations in the Next Decade},
Eds. M.~Matsuoka, N. Kawai, in press, (1997),
astro-ph/9705052.
\bibitem{F98}Fenimore, E. E., {\it ApJ}\ submitted.
\bibitem{FCRS98}Fenimore, E. E.,
Cooper, C., Ramirez, E., and Sumner, M. C.,
{\it ApJ}\ to be submitted.
\bibitem{SF97}Sumner, M. C., and Fenimore, E. E., these proceedings,
astro-ph/9712302.
\bibitem{SP97}Sari, R., and Piran, T., {\it ApJ}\ in press.
\end{references}
\end{document}